\documentclass[12pt,aps,reprint]{revtex4-1}
\usepackage{dcolumn}%
\usepackage{amsmath,amssymb,latexsym}
\usepackage{braket}
\usepackage{xcolor}
\usepackage{graphicx}
\usepackage{hyperref}
\usepackage{siunitx}
\usepackage{multirow}
\usepackage{bm}
\usepackage{cleveref}
\usepackage{sidecap}
\usepackage{capt-of}
\usepackage{geometry}
\usepackage{setspace}
\usepackage{tikz}
\usepackage{wrapfig}

\begin{document}

\title[]{Site-selective-induced isomerization of formamide}
\author{S. Oberli$^1$, J. Gonz\'{a}lez-V\'{a}zquez$^1$, E. Rodr\'{i}guez-Perell\'{o}$^2$, M. Sodupe$^2$, F. Mart\'{i}n$^{1,3,4}$, and A. Pic\'{o}n$^1$}
\affiliation{$^1$Departamento de Qu\'{i}mica, Universidad Aut\'{o}noma de Madrid, 28049 Madrid, Spain}
\affiliation{$^2$Departament de Qu\'{i}mica, Universitat Aut\`{o}noma de Barcelona, 08193 Bellaterra, Spain}
\affiliation{$^3$Instituto Madrile\~{n}o de Estudios Avanzados en Nanociencia (IMDEA-Nanociencia), Cantoblanco, 28049 Madrid, Spain}
\affiliation{$^4$Condensed Matter Physics Center (IFIMAC), Universidad Aut\'{o}noma de Madrid, 28049 Madrid, Spain}

\begin{abstract}
The new capacity of X-ray free-electron laser (XFEL) facilities to produce multi-color X-ray femtosecond pulses paves the way to explore ultrafast phenomena in matter induced by X-ray photons. In the present study, we exploit the site selectivity and the high temporal resolution of a two-color X-ray femtosecond pump-probe sequence to investigate the isomerization of formamide. The pump pulse excites a particular atomic site in the molecule, while the probe pulse captures changes in the chemical environment at a remote atomic site.
\begin{wrapfigure}{r}{0.5\textwidth}
\begin{center}
\includegraphics[width=0.48\textwidth]{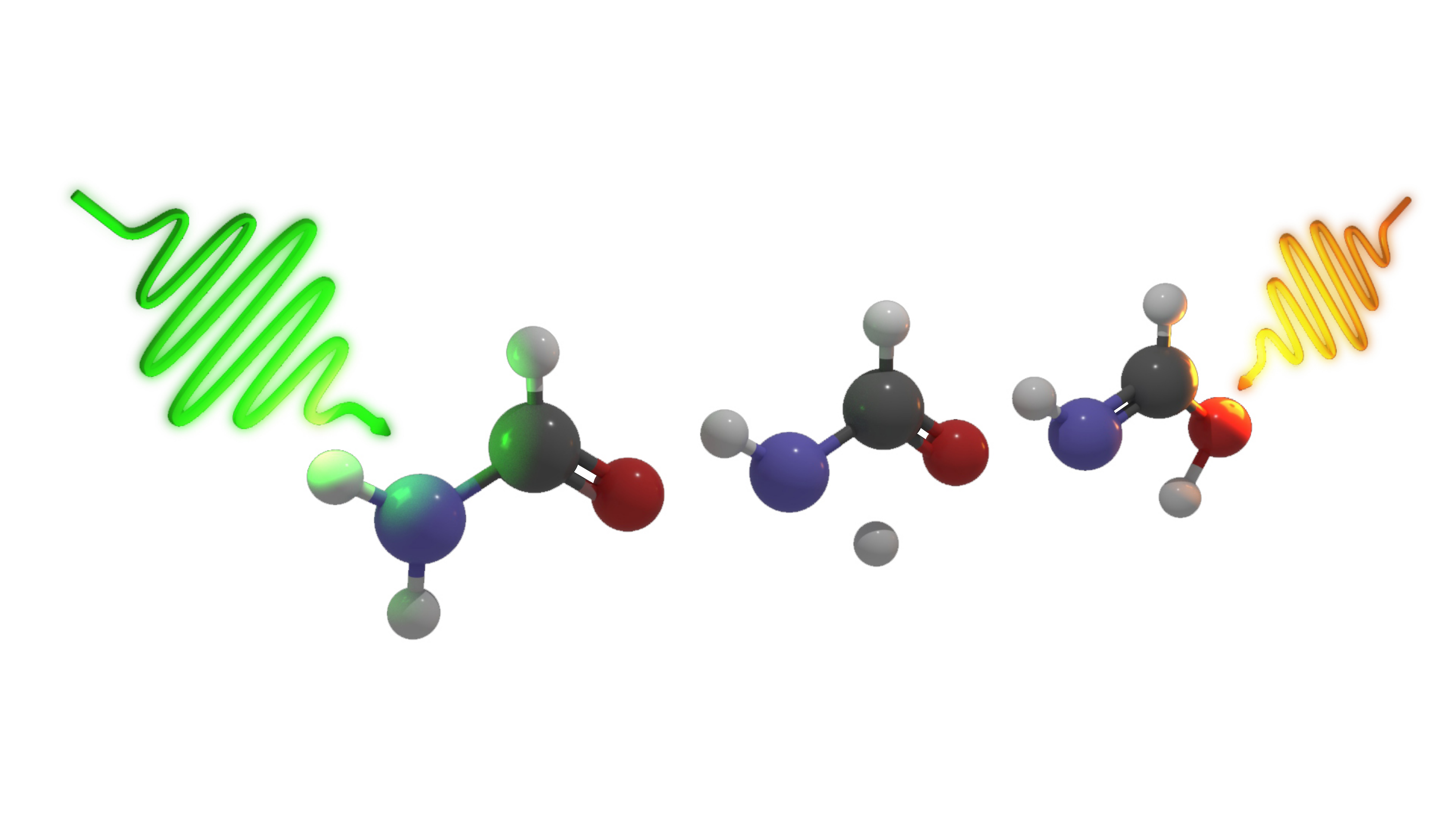}
\end{center}
\end{wrapfigure}
The response of the system is found to strongly depends on the nature of the excited site, and the core excitation provides a selective control on chemical bond breaking. In particular, we show that the $\mathrm{N1s} \rightarrow \pi^*$ transition favors the isomerization reaction in formamide and demonstrate the possibility to observe in real time the hydrogen migration by measuring the time-dependent chemical shifts with the X-ray probe pulse. This work opens a unique perspective for X-ray-induced photochemistry at XFEL facilities.
\end{abstract}
\maketitle

\section{Introduction}
With the revolution of X-ray sources and in particular the advent of X-ray free-electron lasers (XFELs), the production of attosecond/few-femtosecond X-ray pulses opens a new door to the investigation of ultrafast X-ray-induced phenomena in matter. The dynamics is triggered by a high-energy photon that interacts with core electrons that are well localized in the molecule, a completely different regime than in traditional photochemistry.
In this emerging field, some unprecedented applications have already been demonstrated \cite{berrah2017perspective}, such as ultrafast isomerization of acetylene dication \cite{jiang2010ultrafast,liekhus2015ultrafast,li2017ultrafast,osipov2003photoelectron}, hetero-site-specific femtosecond intramolecular dynamics \cite{picon2016hetero} and charge migration \cite{kuleff2016core,picon2018auger,inhester2019detecting}, to name just a few.

In this context, previous studies of X-ray-induced isomerization \cite{jiang2010ultrafast,liekhus2015ultrafast,li2017ultrafast,osipov2003photoelectron} have not yet exploited the site-selective excitation of X rays, which can be an additional knob to trigger and control the reaction in the femtosecond time scale. In this work, we theoretically demonstrate a site-selective-induced isomerization scheme that can be implemented in future XFEL facilities. 

Isomerization plays a key role in nature, for example to trigger the proton transfer by the bacteriorhodopsin protein \cite{mathies1991femtoseconds}. An other molecule of biological interest is formamide, which contains the four most abundant elements in the universe, and is involved in the formation of RNA precursors through UV radiation, nucleobases and other pre-metabolic components \cite{saladino2012formamide,ferus2014high}. Previous studies on formamide focus mainly on the ground state tautomerism \cite{schlegel1982tautomerization} or the effect of solvation on the energetics of proton transfer \cite{wang1991ab,barone1995direct}, while photo-dissociation dynamics of the two isomers was investigated after valence excitation \cite{spinlove2018curve}.
In contrast with the usual photoinduced isomerization reactions involving transitions between delocalized valence molecular orbitals, we propose here a novel investigation of formamide tautomerism in the X-ray regime. 
We take advantage of the unique capabilities of XFELs to produce a two-color X-ray pump-probe sequence \cite{picon2016hetero} to assess the possibility to 1) trigger the isomerization reaction by exciting a core electron of the molecule; 2) influence the hydrogen migration by exciting different atomic sites; 3) observe a fingerprint of the isomerization dynamics through core-ionization of an other atom in the same molecule.
To these purposes, we developed a full quantum model based on highly-correlated electronic structure calculations and a nuclear propagation algorithm to simulate the dynamics in the core-excited states.
We demonstrate the possibility to induce hydrogen migration by taking advantage of site-selective core-excitation. By creating a core vacancy at the C, N, or O atoms, we show that we can selectively weaken and strengthen some bonds in order to hinder or favour hydrogen migration. We observe a clear signature of bond breaking/forming as well as a footprint of the isomerization reaction in the time-dependent X-ray photoelectron spectra. This study paves the way to the investigation of site-selective X-ray-induced ultrafast photochemistry at XFEL facilities.

\section{Hetero-site pump-probe scheme}

\begin{figure*}
\includegraphics{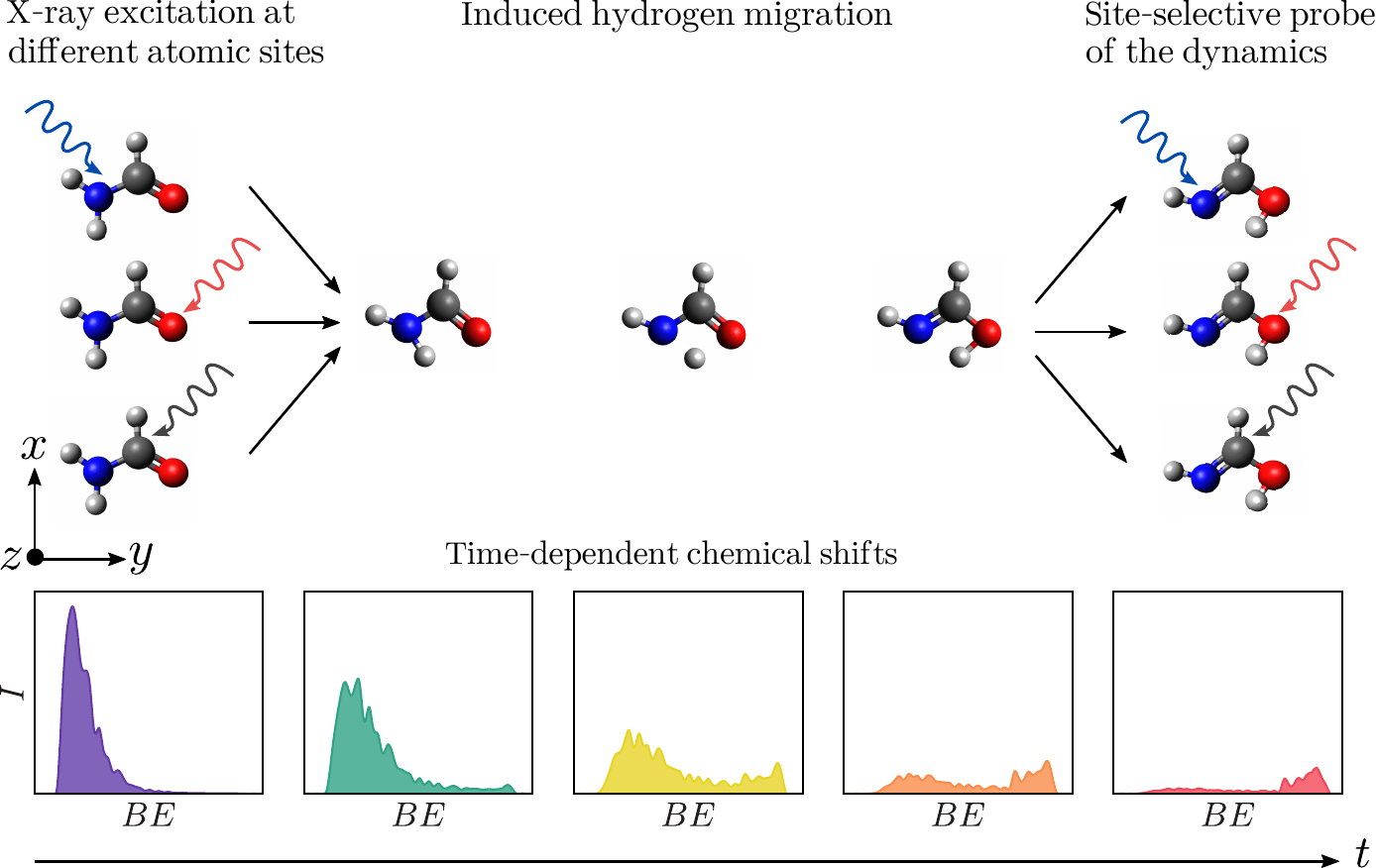}
\caption{\label{fig:geom_opt}Two-color femtosecond X-ray pump-probe scheme to investigate formamide isomerization. a) Different responses of the molecule to core-excitation by the X-ray pump pulse is expected depending on the nature of the excited atom. The dynamics is probed by ionizing the core shell of an other element with an X-ray probe pulse. b) Changes in the local chemical environment of the ionized-atom may be reflected in the binding energy ($BE$) of the core-electron.}
\end{figure*}

The hydrogen migration in formamide is investigated using an hetero-site-specific pump-probe scheme:
A first X-ray pulse initiates the hydrogen migration through core-excitation of a particular site of formamide, while a second X-ray pulse ionizes a core electron either at the C, N or O atom, as illustrated in Figure \ref{fig:geom_opt}.
The measurement of the photoelectron energy with respect to the delay between the two pulses provides information on the changes in the local chemical environment around the absorber atom and thus on structural modifications with femtosecond resolution (see Figure \ref{fig:geom_opt}) \cite{lindgren2004chemical}. This is a well-established technique known as X-ray photoelectron spectroscopy (XPS) \cite{siegbahn1982electron}.
 So far XPS has been mainly used for static spectroscopy but the recent developments in ultrashort-X-ray-pulse sources have led to a renaissance of XPS for time-resolved studies.
 The capacity of XFELs to deliver a sequence of two femtosecond X-ray pulses \cite{lu2018development,lutman2013experimental,marinelli2015high,lutman2016fresh,hara2013two,allaria2013two,ferrari2016widely,penco2018two,bozek2015x} allows for the implementation of the proposed scheme.

\begin{figure*}[t]
\includegraphics{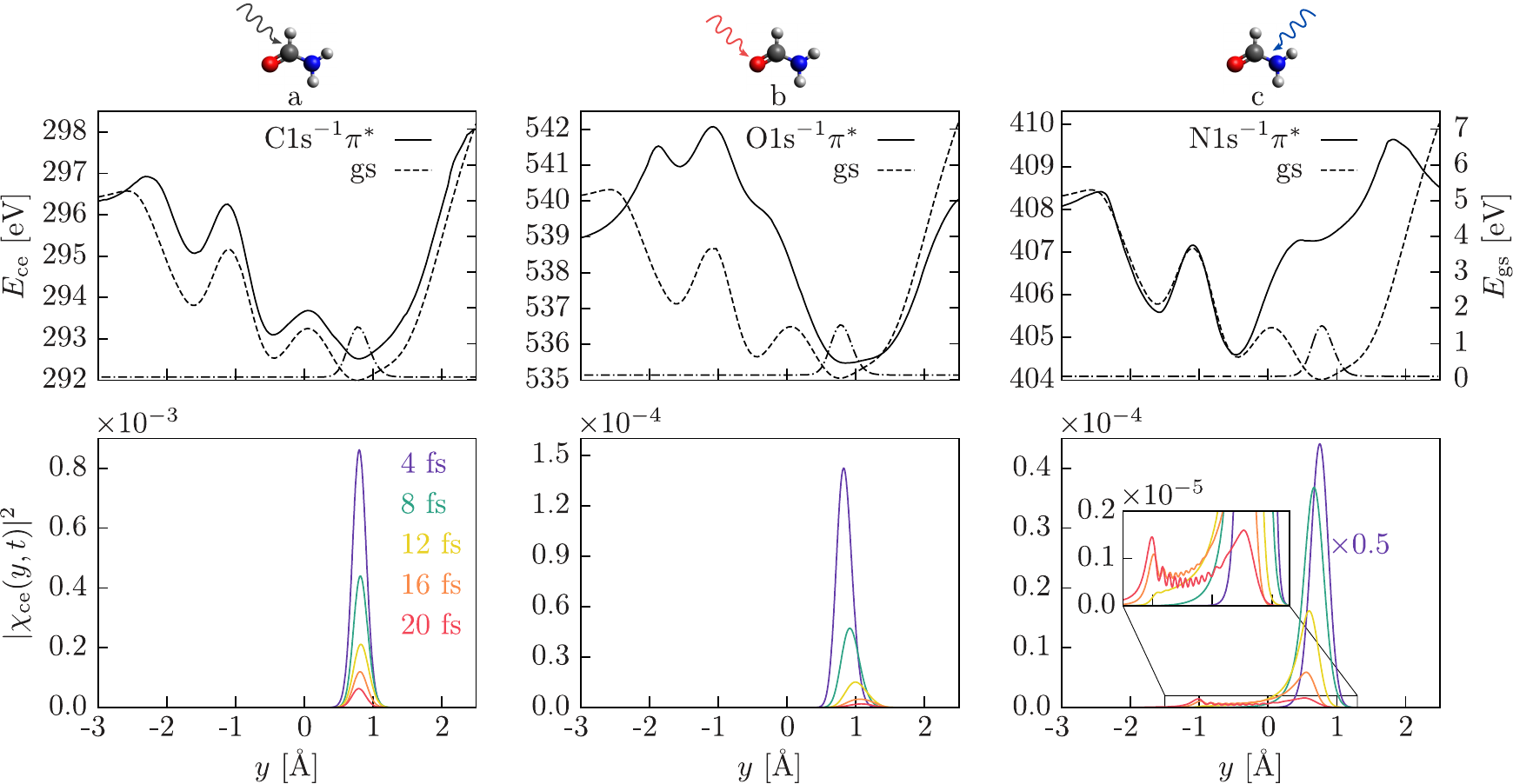}
\caption{\label{fig:PEC_WP}Nuclear wavepacket dynamics in the a) $\mathrm{C1s}^{-1}\pi^*$ b) $\mathrm{O1s}^{-1}\pi^*$ c) $\mathrm{N1s}^{-1}\pi^*$ core-excited states. Upper panel: Comparison of the PECs of the ground state (dashed line) and the core-excited states (solid line). The initial wavepacket in the ground electronic state is also displayed (dash-dotted). Lower panel: The nuclear wavepacket in the core-excited state is shown for some particular times during the propagation.}
\end{figure*}

\section{Site-selective-induced isomerization}

In this section we explore the possibility to trigger and control the isomerization reaction by specifically exciting a particular atom in the molecule (see Figure \ref{fig:geom_opt}).
The excitation energy from the 1s orbital of the C, N and O-sites to the $\pi^*$ orbital is around 292.5, 407.2, and 535.4 eV, respectively. In the simulations we consider X-ray pulses of 1.67-fs full-width at half maximum (FWHM) length, see more details in Section \ref{sec:quantum}. Both electronic and nuclear dynamics are treated in a fully quantum framework in which only the translational motion of the hydrogen (along the $y$-coordinate) is taken into account while the other nuclear degrees of freedom are frozen. This approximation captures the main features of the dynamics under investigation, see more details in Section \ref{sec:1Dmodel}.

We first discuss the response of the molecule when the X-ray pulse interacts with the C site. The PECs of the ground state and the $\mathrm{C1s^{-1}\pi^*}$ core-excited state are shown in the upper panel of Figure \ref{fig:PEC_WP}a. We see that core-excitation of the C has only a small effect on the shape of the PEC, which is very similar to that of the ground state especially in the Franck-Condon region. Therefore the interaction with the pump pulse mainly populates the ground vibrational state of $\mathrm{C1s^{-1}\pi^*}$, i.e. $|\braket{\phi^\mathrm{ce}_0|\phi^\mathrm{gs}_0}|^2=0.99$, $\phi^\mathrm{ce}$ and $\phi^\mathrm{gs}$ being the vibrational states of the core-excited and the ground electronic states, respectively. By exciting the C site the stability of formamidic acid compared to formamide is increased by 0.04~eV and the energy barrier is only lowered by 0.29~eV (see Table \ref{tab:Ebarrier}), such that the nuclear wavepacket remains trapped in the region corresponding to the formamide geometry (see lower panel of Figure \ref{fig:PEC_WP}a). The limited impact of core excitation on the electronic structure is explained by the distances between the carbon atom and those directly involved in the isomerization reaction (the H, N and O atoms). Exciting the core electron in the $\pi^*$ orbital only weakens the delocalized $\pi$-orbitals but does not induce breaking of the N-H bond, such that core excitation of the carbon does not trigger the isomerization reaction.

\begin{table}
\begin{tabular}{ccc}
\hline
Electronic state & $E_\mathrm{f'}-E_\mathrm{f}$~[eV] & $E_\mathrm{ts}-E_\mathrm{f}$~[eV] \\
\hline
gs & 0.62 & 1.45 \\
$\mathrm{C1s^{-1}\pi^*}$ & 0.58 & 1.16 \\
$\mathrm{O1s^{-1}\pi^*}$ & 4.34 & 2.77 \\
$\mathrm{N1s^{-1}\pi^*}$ & -2.68 & -0.97 \\
\hline
\end{tabular}
\caption{\label{tab:Ebarrier}Difference in energy between the formamidic (f') and formamide (f) species, as well as the energy barrier height in the ground and the core-excited states. $E_\mathrm{f}$, $E_\mathrm{f'}$, and $E_\mathrm{ts}$ correspond to the energy of the formamide, formamidic, and transition state at the geometry of the corresponding species in the ground electronic state.}
\end{table}

We consider now the response of the molecule when the X-ray pulse excites the O site. The PECs of the ground and core-excited states as well as the evolution of the nuclear wavepacket in $\mathrm{O1s^{-1}\pi^*}$ are displayed in Figure \ref{fig:PEC_WP}b. In this case there is a large energy barrier of 2.77~eV (see Table \ref{tab:Ebarrier}) in the core-excited state that prevents the isomerization to take place. The equilibrium geometry is displaced by 0.08~\AA~and the nuclear wavepacket evolves towards larger distances, i.e. the N-H bond stretches in the early times of propagation. The formation of formamidic is not favorable in the $\mathrm{O1s^{-1}\pi^*}$ state since the core electron is excited to a delocalized $\pi^*$ orbital, leading to a depletion of the electron density around the oxygen atom which becomes more basic. The strong repulsion between the positively charged oxygen and the hydrogen prevents the isomerization to occur.

The most interesting case is when the X-ray pulse excites the N site, as shown in Figure \ref{fig:PEC_WP}c. The Franck-Condon distribution covers mainly the $18-23^\mathrm{th}$ vibrational states of the core-excited state. We tune the energy of the laser to the resonance corresponding to the population of the $19^\mathrm{th}$ vibrational state that exhibits the largest Franck-Condon factor, and lies above the small energy barrier of $\sim0.02$~eV. The excitation from the 1s to the $\pi^*$ increases the acidity of the $\mathrm{NH_2}$ group, favors the breaking of the CO double bond and also increases the O basicity: As a result the hydrogen migration is triggered as shown in the lower panel of Figure \ref{fig:PEC_WP}c. Around 12~fs the nuclear wavepacket reaches the region around $y=-0.43$~\AA~where the formamidic acid is formed despite the fast Auger decay (5.77~fs) of the core-excited state.  

In summary, the hydrogen migration after the absorption of an X-ray photon strongly depends on the excited atom. This site selectivity provides the possibility to trigger hydrogen migration in the core-excited state and eventually to control the formation of formamidic acid. In the following, we determine the chemical shifts with a second X-ray pulse to investigate these scenarios and show that a clear signature of the hydrogen migration can be obtained. 

\begin{figure*}[ht]
\includegraphics[scale=1]{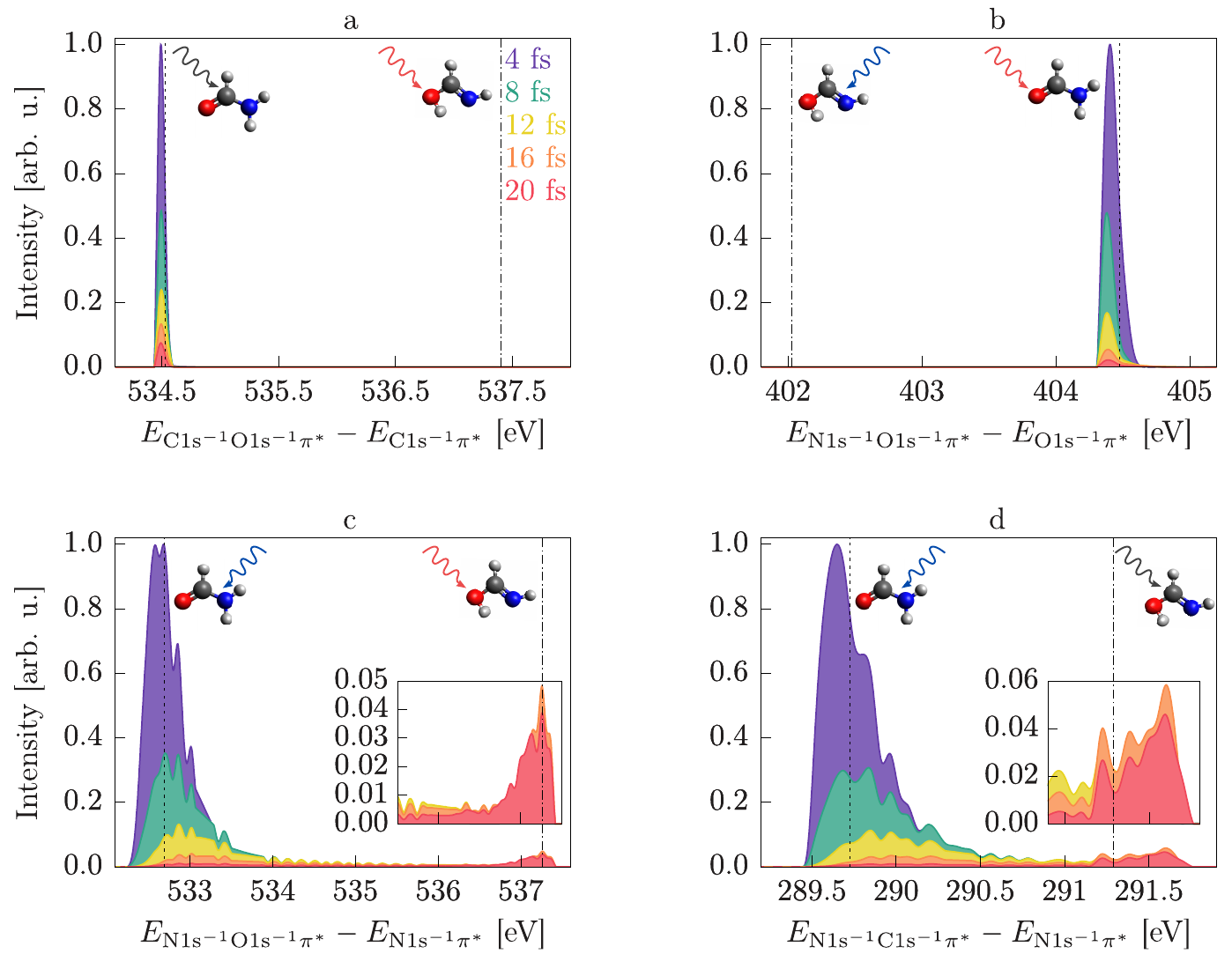}
\caption{\label{fig:TDCS_CNO}Time-dependent chemical shifts associated with core-excitation of one element -- a) C, b) O, c) N, d) N -- followed by core-ionization of a remote atom -- a) O, b) N, c) O, d) C. The vertical dashed (dash-dotted) line corresponds to the chemical shift at the optimized geometry of formamide (formamidic) in the ground electronic state. The intensities are given in arbitrary units.}
\end{figure*}

\section{Ultrafast chemical shifts}

In this section, we use the femtosecond X-ray pump-probe scheme presented in Figure \ref{fig:geom_opt} to calculate time-dependent chemical shifts, i.e. the change in the binding energy of the core electron which is ejected from the molecule by the second pulse as the system evolves in the core-excited state. For a direct comparison with an experimental XPS spectrum, a convolution should be performed to take into account the laser bandwidth ($\sim 1.1$~eV), the finite lifetime of the double-core-hole states produced by the probe pulse ($\hat{\Gamma}_\mathrm{NO}\sim272$~meV and $\hat{\Gamma}_\mathrm{NC}\sim222$~meV), as well as the energy resolution of the photoelectron detector. Details on the calculations can be found in Sections \ref{sec:quantum} and \ref{sec:Electronic}.

The time-dependent chemical shift associated with the C-site excitation followed by the ionization of the O-site is shown in Figure \ref{fig:TDCS_CNO}a. A peak at 534.5~eV, which corresponds to the formamide configuration, appears in the initial times of the dynamics, and the signal losses intensity with respect to time due to the Auger decay. As expected, there is no signal at 537.4~eV which would have been a signature of the formation of formamidic acid. Indeed, the nuclear wavepacket does not have enough energy to overcome the energy barrier that prevents the isomerization to take place in the core-excited state and there are no clear changes in the chemical environment.

The time-dependent chemical shift associated with O-site excitation followed by ionization of the N atom is displayed in Figure \ref{fig:TDCS_CNO}b. As expected from the nuclear dynamics in the $\mathrm{O1s^{-1}\pi^*}$ state, there is no signal at 402 eV in the spectrum since the large energy barrier prevents the hydrogen migration to take place. The positive hole at the O-site has thus no influence on the local chemical environment around the N-site.

The time-dependent chemical shift in which the N-site is excited first followed by the ionization of the O- and C-sites are shown in Figures \ref{fig:TDCS_CNO}c and \ref{fig:TDCS_CNO}d, respectively. In the former case, in the earlier times of the dynamics a peak appears at $\sim532.6$~eV corresponding to the formamide geometry. The intensity of this peak decreases in time due to the Auger decay on one hand, and the evolution of the wavepacket towards a region where the chemical shift is higher on the other hand. Indeed, after 12~fs a signal rises at $\sim537.3$~eV which is a signature that the isomerization reaction has time to occur before the Auger decay completely depopulates the core-excited state. This large variation in the chemical shift ($\Delta E\sim4.7$~eV) is clearly observed and has an intuitive explanation. As the hydrogen approaches the oxygen atom, the local positive charge increases around the oxygen and makes more difficult the ionization of the O-1s core electron, i.e. the binding energy increases. In the case of the C-site, the chemical shift is not as strong ($\Delta E\sim2$~eV): At early times, a peak appears around 289.6~eV when the core-excited state is formed at the formamide geometry, while after 12~fs a small signal is observed around 291.6~eV when the hydrogen is attached to the oxygen.  In summary, the isomerization reaction is barrierless in the $\mathrm{N1s^{-1}\pi^*}$ state and there is a probability that the hydrogen migration occurs despite the fast Auger decay.

\section{Conclusions}

The present study shed light on the ability of femtosecond X-ray pulses to site-selectively induce and unravel the dynamics of the short-lived core-excited molecular states, and opens up promising avenues to design reaction routes inaccessible with standard photochemistry techniques.
In particular, we developed a simple and elegant model to demonstrate the possibility to trigger an isomerization reaction and more generally to control bond breaking/forming in molecules in the X-ray regime. 

Using the prototypical formamide molecule, we show that the response of the system to the excitation of a core electron is very different depending on the element that absorbs the X-ray photon. In the present case, core-excitation of the nitrogen atom triggers the isomerization reaction which is precluded in the ground electronic state due to the energy barrier that separates the formamide from the less stable formamidic species. Despite the fast Auger decay that competes with isomerization in the core-excited state, we can observe a strong shift in the binding energy of the carbon and oxygen core electrons as the hydrogen migration takes place. Instead, core-excitation of the carbon or oxygen atoms does not initiate hydrogen transfer because of the significant energy barrier in these states.

We demonstrate the feasibility to track the hydrogen migration by using time-resolved two-color X-ray photoelectron spectroscopy. This technique is very sensitive to local dynamics and is exploited to detect the bonding of the hydrogen atom to the oxygen in formamide. This work anticipates a unique application that will be available at XFEL facilities with multi-pulse multi-color capabilities.

\section{Theoretical methods}

\subsection{Quantum model} \label{sec:quantum}

The core-excited state dynamics is investigated using a two-color X-ray pump-probe scheme: The pump pulse induces the formation of a core-excited state, whose dynamics is probed by resorting to the time-dependent chemical shifts, treating the probe pulse in first-order perturbation theory. Atomic units are used in the following equations.

The formation of core-excited species induced by the interaction of the molecule with an X-ray pump pulse is described by solving the time-dependent Schr\"{o}dinger equation (TDSE):
\begin{eqnarray}
(\hat{H}+\hat{W}(t))\ket{\Psi(t)}=\mathrm{i}\frac{\mathrm{d}}{\mathrm{d}t}\ket{\Psi(t)},
\label{eq:TDSE}
\end{eqnarray}
with $\hat{H}=\hat{T}_\mathrm{n}+\hat{h}_\mathrm{e}$ the unperturbed Hamiltonian given by the sum of the nuclear kinetic energy $\hat{T}_\mathrm{n}$ and the electronic Hamiltonian $\hat{h}_\mathrm{e}$. The transition dipole operator $\hat{W}(t)$ describes the interaction of the molecule with the pump pulse. The molecular wavepacket is given by the sum for each electronic state $i$ of the product of the nuclear wavepacket $\ket{\chi_i(t)}$ and the electron wavefunction $\ket{i}$:
\begin{eqnarray}
\ket{\Psi(t)}=\ket{\chi_\mathrm{gs}(t)}\otimes\ket{\mathrm{gs}}+\ket{\chi_\mathrm{ce}(t)}\otimes\ket{\mathrm{ce}},
\label{eq:MolWP}
\end{eqnarray}
where $\mathrm{gs}$ and $\mathrm{ce}$ denote the ground and the core-excited states, respectively. A two-state model is a good approximation in the present case, as justified in Section \ref{sec:Electronic}.
In order to derive the equations of motion for the nuclei, we substitute the ansatz for the total molecular wavefunction (\ref{eq:MolWP}) in the TDSE (\ref{eq:TDSE}) and take the scalar product with the different electronic wavefunctions:
\begin{eqnarray}
\mathrm{i}\frac{\mathrm{d}}{\mathrm{d}t}\ket{\chi_\mathrm{gs}(t)}&=&\big(\hat{T}_\mathrm{n}+\hat{V}_\mathrm{gs}\big)\ket{\chi_\mathrm{gs}(t)} \nonumber \\
&&-\braket{\mathrm{gs}|\vec{r}|\mathrm{ce}}\ket{\chi_\mathrm{gs}(t)}\vec{E}(t), \nonumber \\
\mathrm{i}\frac{\mathrm{d}}{\mathrm{d}t}\ket{\chi_\mathrm{ce}(t)}&=&\big(\hat{T}_\mathrm{n}+\hat{V}_\mathrm{ce}-\mathrm{i}\frac{\hat{\Gamma}_\mathrm{ce}}{2}\big)\ket{\chi_\mathrm{ce}(t)} \nonumber \\
&&-\braket{\mathrm{ce}|\vec{r}|\mathrm{gs}}\ket{\chi_\mathrm{ce}(t)}\vec{E}(t).
\label{eq:EOM}
\end{eqnarray}
Non-adiabatic couplings are not included in our model since they are negligible in the region of interest, see Section \ref{sec:Electronic} for more details.
The potential energy curves of the ground and the core-excited states are denoted $\hat{V}_\mathrm{gs}$ and $\hat{V}_\mathrm{ce}$, respectively. 
The Auger decay width $\hat{\Gamma}_\mathrm{ce}$ is inversely proportional to the core-hole lifetime, i.e. $\hat{\Gamma}_\mathrm{ce}=1/\tau_\mathrm{ce}$, and has been introduced phenomenologically to take into account the loss of population due to the Auger decay of the excited species.
The Auger decay widths are taken as constant and are assumed to be identical to the atomic ones for a similar core-excited state. Their values are taken from literature and are summarized in Table \ref{tab:Auger}.
The Auger decay width appears naturally if the Auger states are included in the ansatz (\ref{eq:MolWP}) \cite{cederbaum1981local,oberli2018molecular}.
\begin{table}
\begin{tabular}{ccc}
\hline
Core-excited state & $\hat{\Gamma}_\mathrm{ce}$~[meV] & $\tau_\mathrm{ce}$~[fs] \\
\hline
$\mathrm{C1s^{-1}\pi^*}$ ($\mathrm{CO}$) & 108 & 6.09 \cite{saito2000lifetime} \\
$\mathrm{O1s^{-1}\pi^*}$ ($\mathrm{O_2}$) & 158 & 4.17 \cite{saitoh2001first} \\
$\mathrm{N1s^{-1}\pi^*}$ ($\mathrm{N_2}$) & 114 & 5.77 \cite{saitoh2001first} \\
\hline
\end{tabular}
\caption{\label{tab:Auger}Auger decay widths and Auger lifetimes of the core-excited states with a hole in the carbon, the nitrogen or the oxygen 1s orbital.}
\end{table}

The Gaussian electric field is given by
\begin{eqnarray}
\vec{E}_\mathrm{pump}(t)=E_0\sin{(\omega_0(t-t_0))}e^{-\frac{(t-t_0)^2}{2\sigma^2}}\vec{z},
\label{eq:Laser}
\end{eqnarray}
where the spatial dependence of the pulse envelope is neglected.
$E_0$ is the field strength, $\omega_0$ the central frequency and $t_0$ the temporal center of the pulse. A linearly-polarized laser pulse along the $\vec{z}$-axis is assumed. The pulse duration is given by $t_\mathrm{FWHM}=2\sqrt{\ln{2}}\times\sigma$= 1.67~fs, corresponding to a bandwidth of $\sim1.1$~eV for a Fourier-transform-limited pulse. The intensity of the pulse is $I=10^{14}~\mathrm{W/cm}^2$. The frequency $\omega_0$ of the pump pulse is fixed at the resonance associated with the largest Franck-Condon factor, i.e. $\omega_0=$~292.5, 407.2 and 535.4~eV for $\mathrm{C1s}\rightarrow \pi^*$ ($v^\mathrm{gs}_0\rightarrow v^\mathrm{ce}_0$), $\mathrm{N1s}\rightarrow \pi^*$ ($v^\mathrm{gs}_0\rightarrow v^\mathrm{ce}_{19}$) and $\mathrm{O1s}\rightarrow \pi^*$ ($v^\mathrm{gs}_0\rightarrow v^\mathrm{ce}_0$), respectively, where $v_i^\mathrm{gs}$ and $v_i^\mathrm{ce}$ refer to the $i^\mathrm{th}$ vibrational level of the ground and core-excited electronic states.

In our model, the time-dependent chemical shifts are determined by considering the X-ray probe in first-order perturbation theory. In this approximation, we calculate for each nuclear geometry the binding energy of the core electron which is ejected in the core-excited state. The probability to get one particular chemical shift $E_i$ is obtained from the nuclear distribution in the core-excited state:
\begin{eqnarray}
\rho(E_i,t)dE_i=\int dR |\chi_\mathrm{ce}(R,t)|^2
\label{eq:TDCS}
\end{eqnarray}
with
\begin{eqnarray*}
E_i-\frac{dE_i}{2}<BE_\mathrm{ce}(R)<E_i+\frac{dE_i}{2}
\end{eqnarray*}

where $BE_\mathrm{ce}(R)=E_\mathrm{ce/ci}(R)-E_\mathrm{ce}(R)$ is given by the difference in energy between the core-excited/core-ionized (ce/ci) and the core-excited (ce) states. We select an energy resolution of $dE_i=0.054$~eV. 

\subsection{\label{sec:1Dmodel}Justification of the one-dimensional model for nuclear propagation}

In this section, we assess the accuracy of our one-dimensional model to describe formamide isomerization, where only the $y$-coordinate of the hydrogen atom is considered as active in the dynamics. The geometries of the formamide, the formamidic acid and the transition state in the ground electronic state are optimized at the complete active space self-consistent field CASSCF(12,10) level of theory using a 6-31g basis set with the Molcas quantum chemistry package \cite{karlstrom2003molcas}.

\begin{figure*}
\centering
\includegraphics{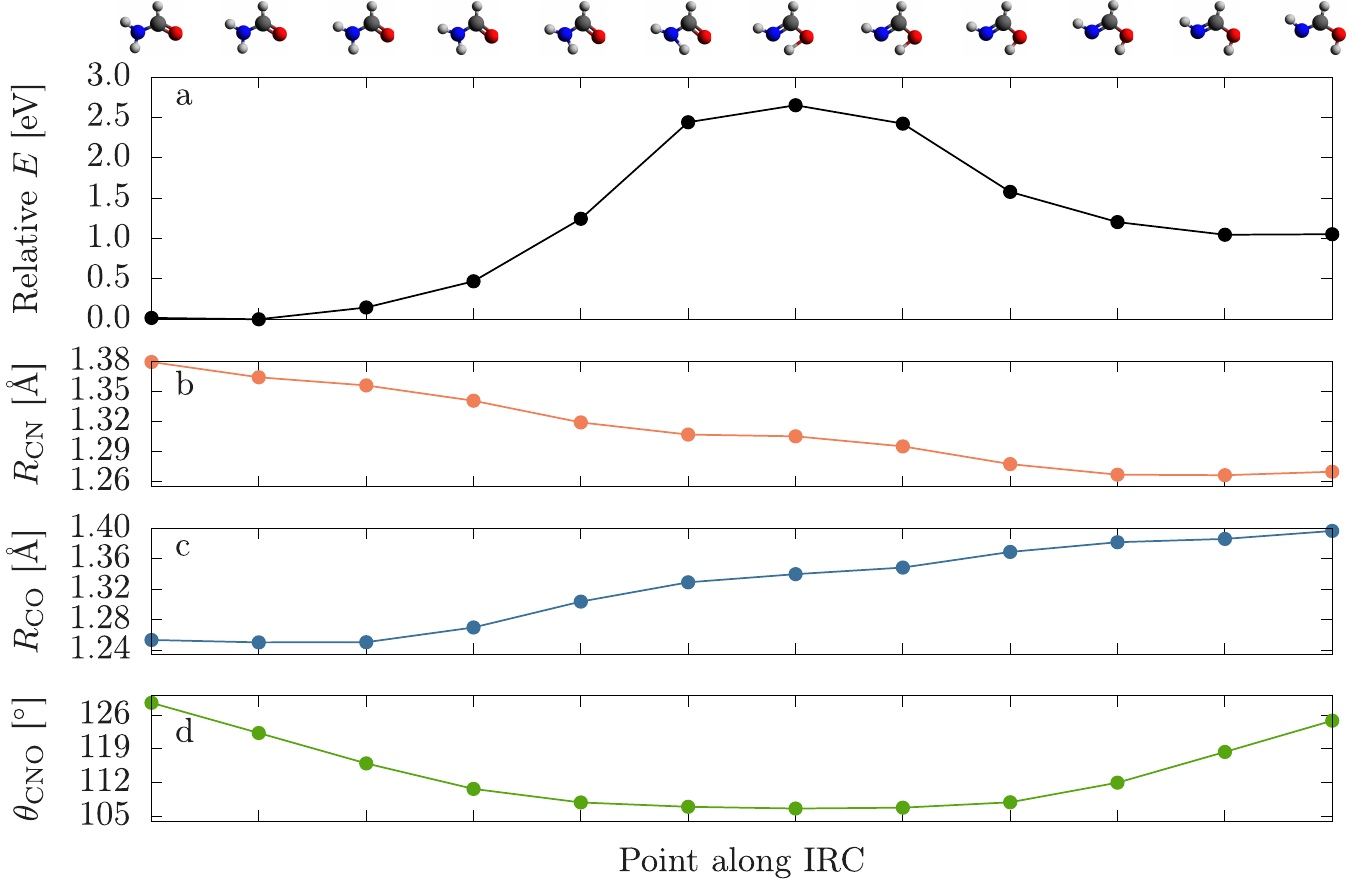}
\caption{\label{fig:IRC}Properties of the molecule in its ground electronic state along the internal reaction coordinate (IRC) in full dimension: a) Relative energy with respect to the formamide species; b) CN internuclear distance $R_\mathrm{CN}$; c) CO internuclear distance $R_\mathrm{CO}$; d) Angle between the C, N and O atoms $\theta_\mathrm{CNO}$.}
\end{figure*}

Starting from the optimized geometry of the transition state, we perform an internal reaction coordinate (IRC) calculation in full dimension at the CASSCF level in Cs symmetry. The changes in energy and some molecular properties (CO and CN bond distances and the angle between C, O and N atoms) along the IRC coordinate are displayed in Figure \ref{fig:IRC}. We see that the molecule is slightly distorted during the hydrogen migration but remains practically planar. 
The energy barrier and the difference in energy between the two isomers for the full-dimensional case (1d-model) amount to 2.64 (1.45)~eV and 1.15 (0.63)~eV, respectively. Although these energies are smaller in the 1d-model, their relative ratio are maintained and the 1d-model grabs the relevant features, describing then the main trends of the dynamics.

Therefore, to a first approximation one considers that the isomerization reaction is mainly driven by the translational motion of the hydrogen between the oxygen and nitrogen atoms (referred to as the $y$-coordinate in the text) which does not differ qualitatively from the real dynamics, while the rest of the molecule remains frozen \cite{hirota1974molecular}.

\subsection{\label{sec:Electronic}Electronic structure calculations}

\begin{figure*}
\centering
\includegraphics{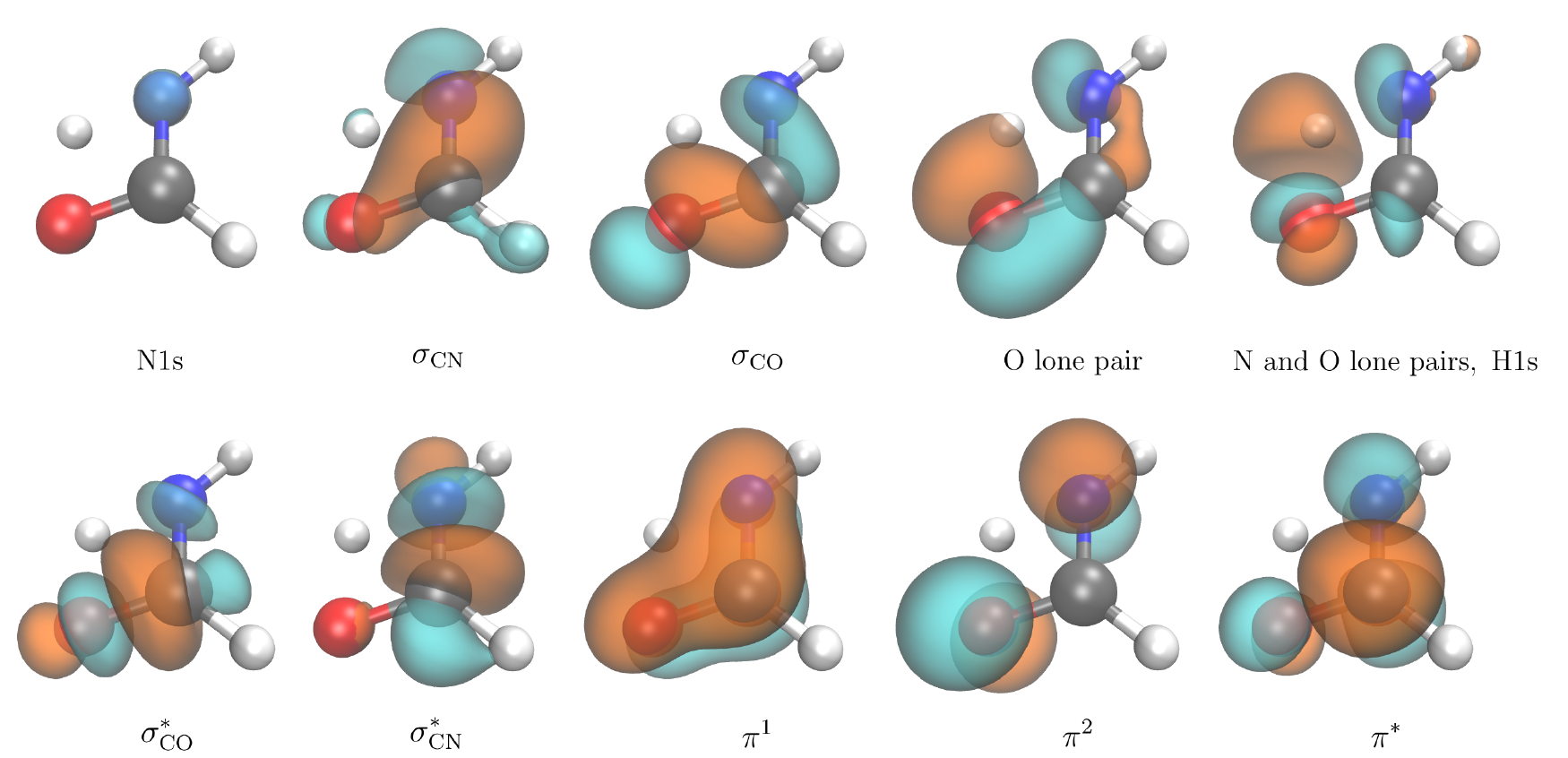}
\caption{\label{fig:MOs}Natural molecular orbitals included in the active space of the ground state and core-excited states, optimized at the CASSCF level of theory.}
\end{figure*}
The potential energy curves (PECs) are obtained from the optimized geometry of the transition state (see Section \ref{sec:1Dmodel}). The PECs are calculated in Cs symmetry using the correlation-consistent polarized valence quadruple-zeta (cc-pVQZ) basis set of Dunning. The active space is prepared at the transition state geometry. For the ground and the core-excited states, we perform a CASSCF(12,10) calculation, where the active space contains 10 orbitals occupied by 12 electrons. Except for the 1s orbital from which the electron is excited, the 6 lowest molecular orbitals (MOs) are closed-shells. Figure \ref{fig:MOs} shows the natural MOs included in the active space, which is composed of 7 MOs of $a^\prime$ symmetry (C1s or N1s or O1s, $\sigma_\mathrm{CN}$, $\sigma_\mathrm{CO}$, $\sigma^*_\mathrm{CO}$, $\sigma^*_\mathrm{CN}$, O lone pair, N lone pair + O lone pair + H1s) and 3 orbitals of $a^{\prime\prime}$ symmetry (two $\pi$ and one $\pi^*$). As the reaction proceeds, the initial N-H $\sigma$ and $\sigma^*$ orbitals together with the O lone pair will evolve to the O-H $\sigma$ and $\sigma^*$ while a lone pair is formed on the N. The delocalized $\pi$ network is affected since the initial C-O double bond breaks to form a C-N double bond. Concerning the core-excited/core-ionized states, we perform a CASSCF(13,11) to additionally include in the active space the 1s orbital of the ionized element. 

The PECs of the ground electronic state is shown in the upper panel of Figure \ref{fig:PEC_WP} (dashed line). The global minimum at $y=0.77$~\AA~corresponds to the formamide species, while the local minimum at $y=-0.43$~\AA~is associated with the formamidic compound. The formamidic acid is less stable by $\sim0.63$~eV than formamide. The transition state is located at $y=0.05$~\AA. An energy barrier in the order of 1.45~eV separates the formamide and formamidic species and prevents the isomerization reaction to occur (see Table \ref{tab:Ebarrier}). There is another local minimum at $-1.6$~\AA~where the hydrogen is at the equilibrium bond distance from the oxygen. For $y>1$~\AA~the energy starts to increase due to the steric interaction between the two hydrogen attached to the nitrogen atom, while for $y<-2.5$~\AA~the energy decreases when the O-H bond dissociates.

The PECs of the core-excited states, where an electron from the 1s orbital of either the C, the O or the N is promoted to the $\pi^*$ orbital are displayed in the upper panel of Figure \ref{fig:PEC_WP} (solid lines). Only singlet states are considered in the calculations since their contributions dominate over the triplet states.
These PECs correspond to the lowest energetic state of a state-average (SA) CASSCF calculation over 15 states. These close-by states arise from shake-up processes, where core-excitation is accompanied by a valence-to-valence excitation \cite{lablanquie2011properties}. Non-adiabatic couplings between the ground state of the core-excited state and the shake-up states can be ignored since they play no role in the region where the dynamics takes place. For the same reason, the contribution of shake-up states are not important during the resonant excitation at the equilibrium geometry.

The PECs associated with core-excited/core-ionized doublet states, required for the XPS spectrum calculations, are also obtained from a SA15-CASSCF calculation.

The TDMs between the ground electronic state and the core-excited states are calculated for each geometry where the dynamics will take place at the CASSCF level.

\section{Acknowledgments}
S. Oberli and A. Pic\'{o}n acknowledge funding from Comunidad de Madrid through TALENTO grant ref. 2017-T1/IND-5432.
F. Mart\'{i}n and J. Gonz\'{a}lez-V\'{a}zquez acknowledge the MINECO projects FIS2016-77889-R. F. Mart\'{i}n acknowledges the Severo Ochoa Programme for Centres of Excellence in R\&D (SEV-2016-0686) and the Mar\'{i}a de Maeztu Programme for Units of Excellence in R\&D (MDM-2014-0377).

\bibliography{references}

\end{document}